\documentclass[times, twoside, watermark]{StyleBioRxiv}
\usepackage{wrapfig}
\makeatletter
\newenvironment{bfig}
 {\def\@captype{figure}}
 {}
\makeatother

\leadauthor{Bo} 

\begin{document}

\title{LabPipe: an extensible informatics platform to streamline management of metabolomics data and metadata}
\shorttitle{LabPipe}

\author[1,2]{Bo Zhao}
\author[3]{Luke Bryant}
\author[1,3]{Michael Wilde}
\author[3]{Rebecca Cordell}
\author[5]{Dahlia Salman}
\author[5]{Dorota Ruszkiewicz}
\author[1]{Wadah Ibrahim}
\author[2]{Amisha Singapuri}
\author[2,3,4]{Tim Coats}
\author[1]{Erol Gaillard}
\author[1]{Caroline Beardsmore}
\author[2,4]{Toru Suzuki}
\author[2,4]{Leong Ng}
\author[2]{Neil Greening}
\author[5]{Paul Thomas}
\author[3]{Paul S. Monks}
\author[1,2]{Christopher Brightling}
\author[1,2]{Salman Siddiqui}
\author[1,2,*]{Robert C. Free}

\affil[1]{Department of Respiratory Sciences, University of Leicester, Leicester, UK}
\affil[2]{NIHR Leicester Biomedical Research Centre, Leicester, UK}
\affil[3]{Department of Chemistry, University of Leicester, Leicester, UK}
\affil[4]{Department of Cardiovascular Sciences, Cardiovascular Research Centre, University of Leicester, Leicester, UK}
\affil[5]{Department of Respiratory Sciences, University of Leicester, Leicester, UK}

\maketitle

\begin{abstract}
Summary:\\
Data management in clinical metabolomics studies is often inadequate. To improve this situation we created LabPipe to provide a guided, customisable
approach to study-specific sample collection. It is driven through a local client which manages the process
and pushes local data to a remote server through an access controlled web API. The platform is able to support data
management for different sampling approaches across multiple sites / studies and is now an essential
study management component for supporting clinical metabolomics locally at the EPSRC/MRC funded
East Midlands Breathomics Pathology Node.\\
Availability and Implementation:\\
LabPipe is freely available to download under a non-commercial open-source license (NPOSL 3.0) along with documentation and installation instructions at http://labpipe.org.\\
\end {abstract}


\begin{corrauthor}
rob.free\at le.ac.uk
\end{corrauthor}

\section*{Introduction}
The nature of multi-disciplinary biomedical studies makes it difficult to efficiently and effectively handle the collation and collection of data sets from different sites and research groups. Effective data management is particularly important in large clinical metabolomics studies, where equipment is distributed across multiple sites with different operating protocols. Moreover, the greater the number of groups and sites involved, the more difficult it becomes to manage this data using manual approaches such as removable storage and paper-based forms.

Software tools are available \citep{Muller2017} which will manage some aspects of the data collection process, however, these do not fully support key parts of data management, which led to the emergence of LabPipe. This was created to streamline data handling, by i) supporting semi-automated sample data collection, linkage and transfer; ii) facilitating collection/data entry of biomedical meta-data linked to the samples; and iii) enabling sample collection notifications. This manuscript describes the LabPipe platform and its successful deployment in a large-scale breathomics project, the East Midlands Breathomics Pathology Node (EMBER) study, handling data from multiple sites and instruments.

\section*{Implementation}

LabPipe was designed using a modular client/server approach (Figure \ref{fig:01}) with an extensible plugin architecture enabling new features to be added when required.

\begin{center}
\begin{bfig}
\includegraphics[width=\linewidth]{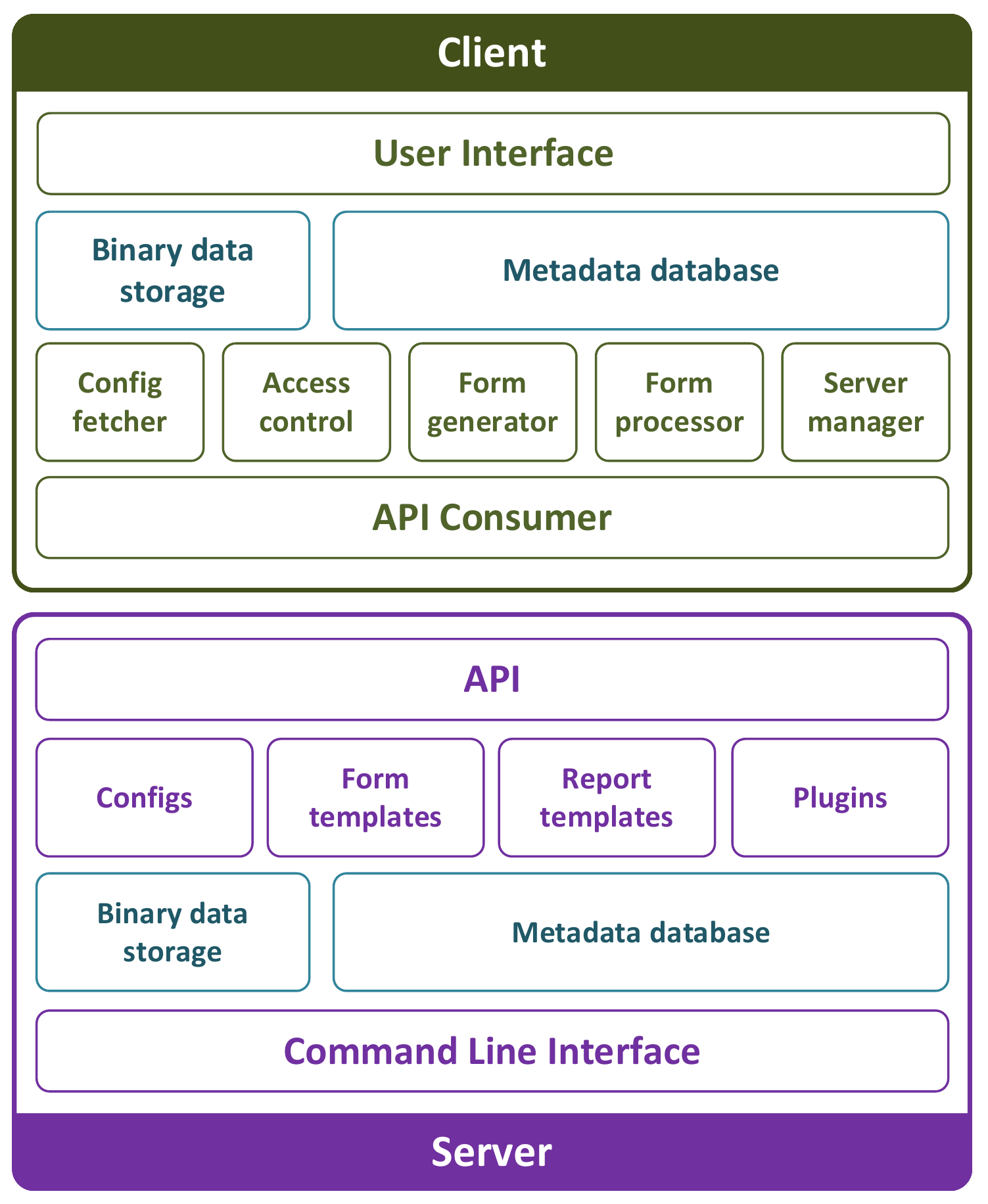}
\caption{Overview of the LabPipe technical architecture.}
\label{fig:01}
\end{bfig}
\end{center}

At the core of LabPipe is LPserver, a light-weight web service-based API. This API provides role-based authorisation to prevent access from non-authorised users. The back-end uses a NoSQL database which enables great flexibility when storing metadata. It allows LabPipe to support different collection configurations (which can be loaded using either the command line or API). These configurations are fetched by each client upon set up and on each use when networked. LPserver also provides web services allowing access to metabolomics data/metadata that has been uploaded.

The LabPipe client tool (LPclient) was developed using the Electron framework, so it can be deployed on multiple operating systems. LPclient is installed locally on computers, providing an integrated front-end to support local experimental data collection and to manage data on the server. LPclient generates forms for different collection protocols using LPserver derived configurations and form templates. Sample metadata can be entered at the sample collection point, LPclient then links this metadata to data files. Additionally, depending on the instrument and operating protocol, client form processes can be customised to handle linkage of data in various ways. For instance, LPclient can generate file IDs to use based on the metadata entered, or identify new/changed files and automatically link these to the metadata.

\section*{Results and Discussion}
\subsection*{EMBER}
The LabPipe tool was developed initially to support the EMBER study \citep{ibrahim_assessment_2019}, a multi-disciplinary breathomics study involving staff from different backgrounds. EMBER involved multiple clinic visits at which clinical staff collected breathomics data and samples using analytical chemistry equipment. Each visit generated multiple sets of metabolomics data which were effectively managed using LabPipe. The tool also allowed pertinent metadata to be entered at the time samples were taken. 

Breath sampling instruments in the EMBER project included both online technologies (in which samples were analysed in real-time on the instrument) and offline technologies (in which samples were collected and later transferred to be analysed in the lab). The approach to data management varied by instrument, processing time and metabolomics software used. In EMBER, LabPipe supported nine instruments (covering four approaches to breath sampling) across two sites.

\subsection*{Research and Collaboration}
Prior to the deployment of LabPipe, data and metadata were handled manually in the EMBER study. This was complex to handle for non-technical staff; required paper-based records; and increased the likelihood of erroneous data re-entry and data loss / poor quality data due to manual collation/linkage of data from individual USB storage devices.

A close collaboration between informaticians, researchers and clinical staff  led to a tool which was accessible and easy to use; guiding non-technical staff through the entire process of sample collection; uploading data to LPserver; and providing data access as painlessly as possible. Development was iterative and helped by qualitative surveys and informal team discussions to assess pain-points in the software - resulting in a better user interface and experience.

The introduction of LabPipe streamlined research by reducing manual data handling and management and saving time and effort. Improved data handling made it easier for researchers to carry out data analysis and link metabolomics data with clinical data. Additionally, the extensible nature of LabPipe meant the tool could evolve to meet the study's needs by adding extensions (e.g. sending researchers notification emails when samples were collected).

Locally installed clients meant LabPipe could integrate with the file system and handle file transfers automatically without requiring manual uploading of data through a web portal. Despite, problems with intermittent wireless network access, a built-in fall-back option meant LPclient would transfer files to LPserver when a network connection became available.

\subsection*{Other options}
Before developing LabPipe, we investigated existing bioinformatics data management tools using the BIBBOX resource as a reference \citep{Muller2017}. Most of the documented tools specialised in particular aspects of data collection and management. For instance, i2b2 \citep{murphy_serving_2010}, tranSMART \citep{bierkens_transmart_2015} and LabKey \citep{nelson_labkey_2011} focused on data warehousing; OpenSpecimen \citep{mcintosh_catissue_2015} tissue tracking; REDCap \citep{harris_research_2009} and OpenClinica \citep{cavelaars_openclinica_2015} electronic Clinical Research Forms (eCRF) data collection; and SeedDMS \citep{USteinmann} document sharing and storing. While these are undoubtedly useful tools, none of them supported our key need to create streamlined data flows with automated transfer of data from local PCs. While there may be commercial proprietary tools which provide some of the features of LabPipe, to the best of our knowledge this side of data management is not handled by existing open/free informatics solutions.

\subsection*{Future work}
Development of LabPipe has focused on metabolomics studies, but the platform is generic enough that it could be used to handle data collection from other lab-based approaches (e.g. flow cytometry, real time PCR etc.). Future work will look at how LabPipe could be adapted to automatically transform/process data into data standards such as ISA \citep{rocca-serra_isa_2010}, for which integrated support for ontologies would be required. To facilitate this, the form handling capabilities are being improved to allow more complex setup such as multi-faceted data entry, enabling eCRF data collection alongside the collection of sample meta-data. Additionally, auditing capabilities are being enhanced so that data access and entry is fully logged.

\section*{Conclusion}
LabPipe provides a user friendly, streamlined, yet fully configurable approach to collecting data and metadata in metabolomics studies. It is now an established component in the Leicester BRC and is supporting multi-disciplinary metabolomics research in Leicester through its underpinning of seven existing studies and similar upcoming research.\\

\section*{Funding}
This work was supported by the Medical Research Council (MRC), Engineering and Physical Sciences Research Council (EPSRC) Stratified Medicine Grant for Molecular Pathology Nodes (Grant No. MR/N005880/1), Midlands Asthma and Allergy Research Association (MAARA) And British Lung Foundation (Grant No. BLFPHD17-1). The work was carried out at the University Hospitals of Leicester NHS Trust, University of Leicester and Loughborough University, supported by the NIHR Leicester Biomedical Research Centre and the NIHR Leicester Clinical Research Facility. The views expressed are those of the author(s) and not necessarily those of the NHS, the NIHR or the Department of Health and Social Care. The authors would like to acknowledge the invaluable efforts of the research nurses responsible for the in-clinic sample collection as well as the input from the wider East Midlands Breathomics Pathology Node consortium (members list can be found at: https://ember.le.ac.uk/web).

\begin{acknowledgements}
The authors would like to thank the EMBER nursing team for working with the LabPipe team and providing feedback and comments.
\end{acknowledgements}

\section*{Bibliography}
\bibliography{labpipe}

\begin{thebibliography}{10}
\providecommand{\natexlab}[1]{#1}
\providecommand{\url}[1]{\texttt{#1}}
\expandafter\ifx\csname urlstyle\endcsname\relax
  \providecommand{\doi}[1]{doi: #1}\else
  \providecommand{\doi}{doi: \begingroup \urlstyle{rm}\Url}\fi

\bibitem[M{\"{u}}ller et~al.(2017)M{\"{u}}ller, Malservet, Quinlan, Reihs,
  Penicaud, Chami, Zatloukal, and Dagher]{Muller2017}
Heimo M{\"{u}}ller, Nicolas Malservet, Philip Quinlan, Robert Reihs, Matthieu
  Penicaud, Antoine Chami, Kurt Zatloukal, and George Dagher.
\newblock {From the evaluation of existing solutions to an all-inclusive
  package for biobanks}.
\newblock \emph{Health and Technology}, 7\penalty0 (1):\penalty0 89--95, mar
  2017.
\newblock ISSN 2190-7196.
\newblock \doi{10.1007/s12553-016-0175-x}.

\bibitem[Ibrahim et~al.(2019)Ibrahim, Wilde, Cordell, Salman, Ruszkiewicz,
  Bryant, Richardson, Free, Zhao, Yousuf, White, Russell, Jones, Patel, Awal,
  Phillips, Fowkes, McNally, Foxon, Bhatt, Peltrini, Singapuri, Hargadon,
  Suzuki, Ng, Gaillard, Beardsmore, Ryanna, Pandya, Coates, Monks, Greening,
  Brightling, Thomas, and Siddiqui]{ibrahim_assessment_2019}
Wadah Ibrahim, Michael Wilde, Rebecca Cordell, Dahlia Salman, Dorota
  Ruszkiewicz, Luke Bryant, Matthew Richardson, Robert~C Free, Bo~Zhao, Ahmed
  Yousuf, Christobelle White, Richard Russell, Sheila Jones, Bharti Patel, Asia
  Awal, Rachael Phillips, Graham Fowkes, Teresa McNally, Clare Foxon, Hetan
  Bhatt, Rosa Peltrini, Amisha Singapuri, Beverley Hargadon, Toru Suzuki,
  Leong~L Ng, Erol Gaillard, Caroline Beardsmore, Kimuli Ryanna, Hitesh Pandya,
  Tim Coates, Paul~S Monks, Neil Greening, Christopher~E Brightling, Paul
  Thomas, and Salman Siddiqui.
\newblock Assessment of breath volatile organic compounds in acute
  cardiorespiratory breathlessness: a protocol describing a prospective
  real-world observational study.
\newblock \emph{BMJ Open}, 9\penalty0 (3):\penalty0 e025486, March 2019.
\newblock \doi{10.1136/bmjopen-2018-025486}.

\bibitem[Murphy et~al.(2010)Murphy, Weber, Mendis, Gainer, Chueh, Churchill,
  and Kohane]{murphy_serving_2010}
S.~N. Murphy, G.~Weber, M.~Mendis, V.~Gainer, H.~C. Chueh, S.~Churchill, and
  I.~Kohane.
\newblock Serving the enterprise and beyond with informatics for integrating
  biology and the bedside (i2b2).
\newblock \emph{Journal of the American Medical Informatics Association},
  17\penalty0 (2):\penalty0 124--130, March 2010.
\newblock ISSN 1067-5027, 1527-974X.
\newblock \doi{10.1136/jamia.2009.000893}.

\bibitem[Bierkens et~al.(2015)Bierkens, van~der Linden, van Bochove, Weistra,
  Fijneman, Azevedo, Boiten, Beliën, and Meijer]{bierkens_transmart_2015}
Mariska Bierkens, Wim van~der Linden, Kees van Bochove, Ward Weistra, Remond
  J.~A. Fijneman, Rita Azevedo, Jan-Willem Boiten, Jeroen Beliën, and
  Gerrit~A. Meijer.
\newblock {tranSMART}.
\newblock \emph{Journal of Clinical Bioinformatics}, 5\penalty0 (1):\penalty0
  S9, May 2015.
\newblock ISSN 2043-9113.
\newblock \doi{10.1186/2043-9113-5-S1-S9}.

\bibitem[Nelson et~al.(2011)Nelson, Piehler, Eckels, Rauch, Bellew, Hussey,
  Ramsay, Nathe, Lum, Krouse, Stearns, Connolly, Skillman, and
  Igra]{nelson_labkey_2011}
Elizabeth~K Nelson, Britt Piehler, Josh Eckels, Adam Rauch, Matthew Bellew,
  Peter Hussey, Sarah Ramsay, Cory Nathe, Karl Lum, Kevin Krouse, David
  Stearns, Brian Connolly, Tom Skillman, and Mark Igra.
\newblock {LabKey} {Server}: {An} open source platform for scientific data
  integration, analysis and collaboration.
\newblock \emph{BMC Bioinformatics}, 12\penalty0 (1):\penalty0 71, December
  2011.
\newblock ISSN 1471-2105.
\newblock \doi{10.1186/1471-2105-12-71}.

\bibitem[McIntosh et~al.(2015)McIntosh, Sharma, Mulvihill, Gupta, Juehne,
  George, Khot, Kaushal, Watson, and Nagarajan]{mcintosh_catissue_2015}
Leslie~D. McIntosh, Mukesh~K. Sharma, David Mulvihill, Snehil Gupta, Anthony
  Juehne, Bijoy George, Suhas~B. Khot, Atul Kaushal, Mark~A. Watson, and Rakesh
  Nagarajan.
\newblock {caTissue} {Suite} to {OpenSpecimen}: {Developing} an extensible,
  open source, web-based biobanking management system.
\newblock \emph{Journal of Biomedical Informatics}, 57:\penalty0 456--464,
  October 2015.
\newblock ISSN 15320464.
\newblock \doi{10.1016/j.jbi.2015.08.020}.

\bibitem[Harris et~al.(2009)Harris, Taylor, Thielke, Payne, Gonzalez, and
  Conde]{harris_research_2009}
Paul~A. Harris, Robert Taylor, Robert Thielke, Jonathon Payne, Nathaniel
  Gonzalez, and Jose~G. Conde.
\newblock Research electronic data capture ({REDCap})—{A} metadata-driven
  methodology and workflow process for providing translational research
  informatics support.
\newblock \emph{Journal of Biomedical Informatics}, 42\penalty0 (2):\penalty0
  377--381, April 2009.
\newblock ISSN 15320464.
\newblock \doi{10.1016/j.jbi.2008.08.010}.

\bibitem[Cavelaars et~al.(2015)Cavelaars, Rousseau, Parlayan, de~Ridder,
  Verburg, Ross, Visser, Rotte, Azevedo, Boiten, Meijer, Belien, and
  Verheul]{cavelaars_openclinica_2015}
Marinel Cavelaars, Jacob Rousseau, Cuneyt Parlayan, Sander de~Ridder, Annemarie
  Verburg, Ruud Ross, Gerben~Rienk Visser, Annelies Rotte, Rita Azevedo,
  Jan-Willem Boiten, Gerrit~A. Meijer, Jeroen A.~M. Belien, and Henk Verheul.
\newblock {OpenClinica}.
\newblock \emph{Journal of Clinical Bioinformatics}, 5\penalty0 (1):\penalty0
  S2, May 2015.
\newblock ISSN 2043-9113.
\newblock \doi{10.1186/2043-9113-5-S1-S2}.

\bibitem[Steinmann(2019)]{USteinmann}
Uwe Steinmann.
\newblock {SeedDMS}, 2019.

\bibitem[Rocca-Serra et~al.(2010)Rocca-Serra, Brandizi, Maguire, Sklyar,
  Taylor, Begley, Field, Harris, Hide, Hofmann, Neumann, Sterk, Tong, and
  Sansone]{rocca-serra_isa_2010}
P.~Rocca-Serra, M.~Brandizi, E.~Maguire, N.~Sklyar, C.~Taylor, K.~Begley,
  D.~Field, S.~Harris, W.~Hide, O.~Hofmann, S.~Neumann, P.~Sterk, W.~Tong, and
  S.-A. Sansone.
\newblock {ISA} software suite: supporting standards-compliant experimental
  annotation and enabling curation at the community level.
\newblock \emph{Bioinformatics}, 26\penalty0 (18):\penalty0 2354--2356,
  September 2010.
\newblock ISSN 1367-4803, 1460-2059.
\newblock \doi{10.1093/bioinformatics/btq415}.

\end{thebibliography}

\end{document}